# Using Data Redundancy Techniques to Detect and Correct Errors in Logical Data

Ahmed Sharuvan[1] and Ahmed Naufal Abdul Hadee[2]

[1]Maldives National Defense Force, Malé, Maldives, [2]National Cyber Security Agency, Malé, Maldives

**ABSTRACT** Data redundancy techniques have been tested in several different applications to provide fault tolerance and performance gains. The use of these techniques is mostly seen at the hardware, device driver, or file system level. In practice, the use of data integrity techniques with logical data has largely been limited to verifying the integrity of transferred files using cryptographic hashes. In this paper, we study the RAID scheme used with disk arrays and adapt it for use with logical data. An implementation for such a system is devised in theory and implemented in software, providing the specifications for the procedures and file formats used. Rigorous experimentation is conducted to test the effectiveness of the developed system for multiple use cases. With computer-generated benchmarks and simulated experiments, the system demonstrates robust performance in recovering arbitrary faults in large archive files only using a small fraction of redundant data. This was achieved by leveraging computing power for the process of data recovery.

## I. INTRODUCTION

Data reliability is an issue prevalent in storage systems [1]. Mechanisms for mitigating data corruption are therefore implemented in hardware systems design. This includes the use of error-correction codes (ECC) to fix sector read errors on the fly by the disk controller of a hard disk drive. The use of these techniques effectively increases the reliability of the drives in terms of data integrity. However, not all data corruption incidents can be fixed using the ECC data alone. Occurrences of silent data corruption have been detected in hard disk drives in a study of a large-scale storage system [2]. To alleviate the shortcomings of these basic methods of error correction, techniques such as Redundant Arrays of Inexpensive Disks (RAID) systems have been used to catch integrity errors caused by the failure of the whole storage device or malfunction in an internal component [3], [4].

Advanced systems for data error correction and fault tolerance such as RAID are primarily used by enterprise systems. A personal computer owner with a single hard disk drive does not have the option to use this technology. In this paper, we propose an adaptation of the RAID system for logical data that can be used by the personal computer owner as well to protect and extend the lifetime of their data.

Existing literature and technology on fault-tolerant storage systems focuses on hardware devices and their layouts or a software replication of such systems. The discussed redundancy system and presented methods of data recovery can be generalized and used in the design of any information system that can use a logical error correction scheme regardless of the underlying storage architecture. The primary contributions of this paper can be summarized as follows:

- We study the RAID system and produce an adaptation of it for logical data.
- We devise a theoretical implementation for the discussed adaptation.
- We implement the theoretical model in software, providing a specification of procedures and file formats used.
- We conduct experimentations for the implemented system to test its effectiveness for different use cases.

The rest of this paper is organized as follows: Section II presents works related to using RAID-like systems with logical data. Section III discusses an adaptation of the RAID system for logical data. Section IV proposes a novel theoretical and practical implementation of a RAID-like system for logical data. Section V presents experimentations with the proposed system and its results. In section VI, the paper is concluded by discussing the implications of the research.

## II. RELATED WORK

Wehlus et al. [5] proposed Parity Volume Set Specification v1.0, Parchive in 2001, a RAID-like system that can recover missing files in a set of files. The implementation is very similar to how the RAID system works, despite being a file format specification to be used on logical data. The method of generating parity data imitates the way it is done in disk arrays. Where data blocks are derived from individual disks, Parchive fetches data from each individual file in a file set. This implies an inefficiency where the data files are of different sizes because the parity data must match the size of the largest data file.

In 2002, Nahas et al. [6] published version 2.0 of the Parchive specification. This version addresses the issue of the need to use large chunks of null-byte padding by splitting each data file into fixed-size blocks. It remains as the current version in use.

The core idea with Parchive is such that where an *n* number of data blocks and *m* number of parity blocks are used, the data can be reconstructed using an *n* number of blocks from the data blocks and parity blocks combined.

## III. ADAPTING RAID FOR LOGICAL DATA
### A. A TECHNICAL REVIEW OF RAID

With the assumptions made about disk arrays by disk manufacturers, the time to failure is exponentially distributed, and failures are independent. The Mean Time To Failure (MTTF) of a disk array is given by:

$$\text{MTTF of a Disk Array} = \frac{\text{MTTF of A Single Disk}}{\text{Number of Disks in the Array}} \quad [3] \quad (1)$$

The disk array gets more unreliable with every addition of a disk, reducing the lifetime of the system. To increase the overall reliability, extra disks are used to store redundant data. The RAID system organizes the disks into reliability groups, each with a number of data disks and check disks. Different layouts can be used depending on the requirements. The MTTF of a RAID system is given by:

$$\text{MTTF}_{\text{RAID}} = \frac{(\text{MTTF}_{\text{Disk}})^2}{(D + C \times n_G) \times (G + C - 1) \times \text{MTTR}} \quad [3] \quad (2)$$

Where $D$ is the total number of disks with data, $C$ is the number of check disks in a group, $n_G$ is the number of groups, $n$ is the total number of disks, $G$ is the number of data disks in a group and MTTR is mean time to repair for a single disk. The equation holds for all levels of RAID organizations except for level 6 with double failure recovery capability. The reliability is impacted by the number of check disks, $C$ used in each level.

The redundant organizations start at level 1, where a check disk of mirrored data is used for each data disk, effectively halving the storage efficiency. Hamming [7] codes have been used in Error Correction Code (ECC) memory to increase the reliability of primary memory via the use of extra chips to store redundant data. Level 2 RAID uses this method of bit interleaving within the disk groups to achieve a more cost-efficient method of data redundancy. However, the use of bit interleaving with this method of redundancy means that access to data from less than $G$ disks requires access to the entire group, limiting the rate of I/O. Since disk controllers can identify by themselves when a disk has failed, we can remove the extra check disks used to identify the failed disk and rely on that information when re-assembling data.

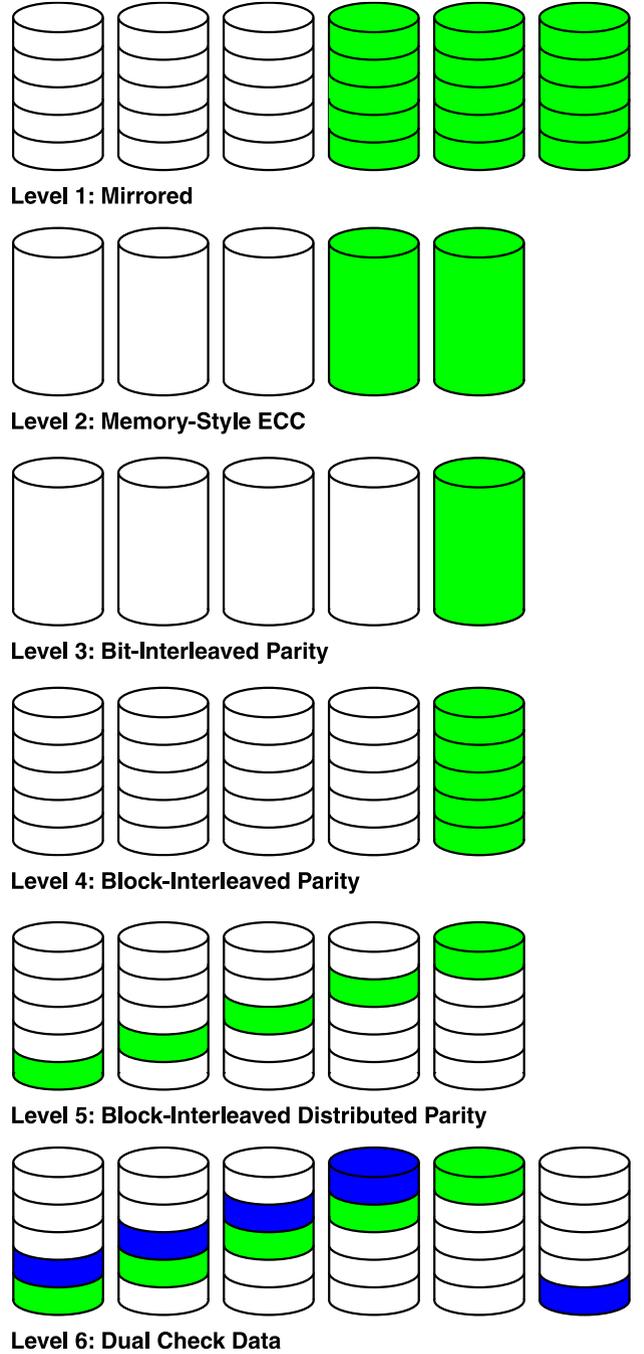

**Level 1: Mirrored**

**Level 2: Memory-Style ECC**

**Level 3: Bit-Interleaved Parity**

**Level 4: Block-Interleaved Parity**

**Level 5: Block-Interleaved Distributed Parity**

**Level 6: Dual Check Data**

**FIGURE 1.** RAID organization levels. Multiple platters on a disk depict block level striping. Colored regions store redundant data.

This enables us to use only one check disk per group, giving the lowest reliability overhead. Park and Balasubramanian [8] proposed a third-level RAID system utilizing these principles. Block level interleaving could instead be used to pair the data as demonstrated by Salem and Garcia-Molina [9] in their implementation of a level 4 RAID system. Transfer units are placed inside the sectors of individual disks. This means that reads can be performed in parallel, increasing the rate of I/O in addition to leveraging the maximum data rate of the disks. However, using a separate disk to keep check information creates a bottleneck for write operations. Parallel write operations are not possible because the check disk needs to be written to whenever updating the data in a data disk. To address this issue, level 5 RAID distributes the check data across all data disks. Now every data disk has a block with parity information. Doing this removes the bottleneck and provides much higher throughput for write operations as well. Albeit level 5 RAID with one check disk per group may be sufficient in terms of the reliability provided, more advanced error correction codes such as Reed-Solomon [10] code can be used to produce an additional set of redundant data, providing a reliability factor of several magnitude higher. Level 6 RAID implements at least two sets of redundant data distributed across the data disks. The formula to get the MTTF of RAID level 6 is:

$$\text{MTTF}_{\text{RAID}} = \frac{(\text{MTTF}_{\text{Disk}})^3}{n \times (G-1) \times (G-2) \times \text{MTTR}^2} \quad [4] \quad (3)$$

Fig. 1 presents an illustration of RAID implementation at each of these levels. The layouts and techniques discussed constitute the design of a hardware RAID system consisting of hard disk drives grouped together. When a RAID-like system is concerned with logical data, elements such as the disk controller are not available and alternative schemes must be developed.

*B. ADAPTATION*

A contiguous blob of data is considered to perform data redundancy on, in turn, to increase the reliability in case of an integrity error. When a parity-based method of redundancy is considered such as XOR parity or Reed-Solomon code, the algorithms consume equal size blocks of data to generate a corresponding block of parity data. In disk arrays, it is easy to distribute these abstract blocks equally among the disks in a group since all disks in a group have the same capacity. The blob of data could be split logically to form several equal-sized blocks. It is possible that the length of the blob is not divisible by any number less than the length itself, in which case, the largest divisible size could be considered for producing redundancy.

Now with the data blob split into several abstract blocks, we can generate redundant data using block interleaved parity. However, when an odd parity is detected at a certain index of a block, we do not have any mechanism to identify the affected block. Moreover, two-bit errors at the same index in different blocks could even erroneously produce the correct output, making the error undetectable using the parity data alone.

Another technique for assuring data integrity, checksum calculation can be used in addition to parity calculation to test the integrity of a given block of data. If we want to simulate the way RAID works, a strong checksum could be assigned to each abstract data block so that we can identify when a block has failed. If the number of occurrences of erroneous blocks does not exceed the number of parity blocks generated, we can successfully reconstruct the data using the good data blocks and parity blocks.

Considering a data blob, no locality of error within an individual abstracted data block is expected since the underlying storage medium is the same. Therefore, it is not necessary that checksums must be computed on a per-block basis. A more extensive use of checksums can be used in combination with parity to increase the reliability further.

Two layers of abstraction can be utilized, referred to as parity blocks and checksum blocks. The checksum blocks are laid out inside every parity block. Error detection sequence can start from computing the checksum blocks. When a checksum mismatch is found, parity errors can be identified within the checksum block contents. It is not necessary that all identified bits with parity errors must be due to a fault in the particular checksum block with the mismatch. This means that combinations can be generated with the identified parity bit indexes and tried until a checksum match is found, giving us the opportunity to incorporate computing power into the process of data recovery. An appropriate checksum output size can be selected so that a computationally feasible number of combinations can give a low enough chance of checksum collision. Also do note that the combination set size given by $2^n - 1$ where $n$ is the number of bit errors grows exponentially. Due to this, it is also important to select an appropriate checksum block size.

**IV. IMPLEMENTATION**
*A. THEORETICAL IMPLEMENTATION*

In this section, a redundancy system is modeled for a data blob, discussing the design and configurations to use, building on the principles from the past section. A mathematical formula will be derived to calculate the system's reliability, which we will use to predict the reliability provided in certain scenarios.

A data blob of arbitrary size is to be split into $p$ blocks of equal length to generate parity data, where $p$ is determined by the redundancy percentage to use. As discussed before, the blob byte length $l$ is more likely not divisible by this number. $l$ divided by $p$ floored, given by $\lfloor l/p \rfloor$ can instead be used as the block length to operate on, neglecting a few extra bytes. In practice, the margin should be minor where the blob size $l$ is significantly larger than $p$.

Now going to the second layer of abstraction, checksum blocks. As per the principles, the checksum blocks must be laid out inside each of the parity blocks. Different layouts are possible. The implementation focuses on performance, therefore we organize the checksum blocks sequentially within each parity block, assuring maximum data read rate during computation. The parity block lengths are selected as a proportion of the blob size. On the contrary, checksum block sizes do not need to be proportional to the blob size. This should instead be determined based on the maximum number of bit errors to be recovered within a data field. Again, the checksum block length may not be divisible by the parity block length. In this case, we simply wrap the last checksum block at the parity block cut-off, giving one block with less than the set size.

There exist several methods for generating parity data. Reed-Solomon code [10], Low-density parity-check code [11] and Turbo code [12] are a few of the available error correction code algorithms in addition to the basic method of XOR parity. In our implementation, we emphasize performance and simplicity, therefore XOR parity is used due to its simplicity and computational efficiency. A limitation of this method however is that only one set of redundant data can be generated.

Just as with parity calculation, we have several available algorithms to calculate checksum. Fletcher [13], Adler [14] and CRC [15] are some of the arithmetic and polynomial algorithms. Fletcher's arithmetic checksum has been demonstrated to provide similar reliability to more complex CRC while being much faster due to its simple and computationally efficient algorithm. This is in line with our principles of performance and simplicity for this implementation. Therefore, we will use a version of Fletcher's algorithm. The different versions Fletcher-8, Fletcher-16, etc. are characterized by their output size. We will now consider an appropriate version to use that gives us a low checksum collision probability when a computationally feasible number of combinations are tried. To do this we can formulate an equation that divides the combination set size by the checksum's output space size given by:

$$w = \frac{2^n - 1}{2^q} \quad (4)$$

Where $w$ is the probability of collision, $n$ is the number of bit errors and $q$ is the checksum output size in bits. In Table 1, we test different parameters with this equation. Do note that the actual collision probability is affected by the properties of the checksum algorithm used. If we consider Fletcher-8, to correct 5-bit errors, we have to run 31 combinations, giving a high collision probability of 12%. This does not even allow us to utilize computing power much. If we consider Fletcher-32, it can correct 26-bit errors by running 67 million combinations, giving a collision probability of 1.56%. We get more error correction capability per checksum data used, however, running this number of combinations can take an inconveniently long amount of time for most current-generation computers. Now when we consider Fletcher-16, it can correct 10-bit errors running 1023 combinations, giving a collision probability of 1.56%. It can correct a good number of bit errors per checksum data used with a set of combinations most computers will test quickly while providing a low collision probability. Therefore, we will use Fletcher-16 in our implementation.

TABLE I
CHECKSUM COLLISION PROBABILITY

| Bit Errors ($n$) | Checksum Output Size ($q$) | Combination Set Size | Collision Probability ($w$) |
|---|---|---|---|
| 5 | 8 | 31 | 0.12109375 |
| 5 | 16 | 31 | 0.0004730224609375 |
| 10 | 16 | 1023 | 0.0156097412109375 |
| 15 | 16 | 32767 | 0.4999847412109375 |
| 15 | 32 | 32767 | 0.0000076291617006 |
| 20 | 32 | 1048575 | 0.0002441403921694 |
| 25 | 32 | 33554431 | 0.0078124997671694 |
| 26 | 32 | 67108863 | 0.0156249997671694 |
| 30 | 32 | 1073741823 | 0.2499999997671694 |

The system consists of two processes, generating redundancy and data recovery. The redundant data will consist of two components, checksum data and parity data. Even though the first layer of abstraction are the parity blocks, we read and write checksum data the first thing in the operations. Therefore, we place checksum data before parity data.

In the redundancy generation process, we start by computing the checksums for each of the checksum block within every parity block sequentially, writing the outputs into a separate redundant data blob. This will give us $qc$ bits of checksum data where $c$ is the number of checksum blocks. Both the number of parity blocks and checksum blocks used should be configurable by the user.

The second step is equally straightforward. We take a bit from each of the parity blocks sequentially, perform XOR operation over each and insert the resulting bit into the redundant data blob. The resulting parity data size should be equal to the size of each parity block $d/p$ bits where $d$ is the number of data bits and $p$ is the number of parity blocks. The redundant data blob should have checksum data and parity data aligned in the respective order.

Once we have the redundant data blob generated, we should be able to use it to recover our original data blob in the case of a loss of integrity. The recovery process starts by iterating the checksum blocks sequentially, calculating and comparing them with the saved checksums. When a mismatch is found, we compute and compare parity data for the checksum block contents, noting the bit indexes with errors. Next, we generate a set of combinations with the erroneous bit indexes, in the order of more values to less values. This is deducing that the current checksum block with error is more likely to cause parity inconsistencies. Bits are flipped with different combinations and calculated to find a matching checksum.

When a match is found, we save the correct combination information in memory and proceed to the next checksum block. If the entirety of possible combinations is tried and no checksum match is found, we note the checksum block index as a failed block. Once the traversal is completed, we go through each of the corrections saved in memory and write the changes if the corrected checksum block index is not marked as a failed block index during the procedure. This is because failed checksum blocks are strongly correlated with erroneous corrections to their pairing blocks.

Limitations of this implementation should be discussed. Table 1 shows that with Fletcher-16, we should be able to reliably correct up to 10-bit errors within a checksum block. If more bit errors are found, both the computation complexity and collision probability will exceed the reasonable limits. Checksum blocks of smaller sizes can be used to prevent the occurrence of this. Secondly, the occurrence of more than one-bit error in a single parity block index can falsely produce the correct parity. This will prevent the error from being corrected. To avoid the occurrence of this, fewer parity blocks can be used. In addition, no integrity loss is expected in the redundant data blob. Therefore, it is advised to store this in a high-reliability environment.

To calculate the theoretical reliability, we can start by finding the probability for checksum collision and parity collision for a given number of $n$ errors. By substitution, we can get the checksum collision probability for the average number of errors in a checksum block and its pairing blocks $(np)/c$, assuming a uniform distribution of errors. Then multiply by $n$ to get the probability for an $n$ number of errors.

$$\Pr(\text{checksum collision}) = \frac{2^{\frac{np}{c}} - 1}{2^q} n \quad (5)$$

Subsequently, parity collision probability can be calculated by finding the cumulative product of the odds that no error bits occur at the same block index by selecting the $i$ value incrementally from 0 to $n$-1 and subtracting this from 1. This model simulates the system by considering a sequence of error bits placed in the data space.

$$\Pr(\text{parity collision}) = 1 - \prod_{i=0}^{n-1}\left(1 - \frac{i}{\frac{d}{p}}\right) \quad (6)$$

Finally, the reliability of the system can be expressed as the probability of successful recovery, given by:

$$\Pr(\text{recovery}) = 1 - \Pr(\text{checksum collision}) - \Pr(\text{parity collision}) \quad (7)$$

$$\Pr(\text{recovery}) = 1 - \left[\frac{2^{\frac{np}{c}} - 1}{2^q} n\right] - \left[1 - \prod_{i=0}^{n-1}\left(1 - \frac{i}{\frac{d}{p}}\right)\right] \quad (8)$$

In Table 2, we use (8) to determine the theoretical reliability provided using different parameters for the variables. It should be noted that the derived figure is an overestimate due to inefficiencies in the checksum algorithm and non-uniformity in the distribution of errors. Configurations are tested with constant values of one megabyte of data bits, 16-bit checksum output size, and 1000-bit errors with varying numbers of parity blocks and checksum blocks. Parity block counts used are to make 50%, 20%, 10%, and 5% of data blob size, in the corresponding order. Checksum block counts used are to make 64 bytes and 128 bytes block size where checksum data amounting to 3.125% and 1.5625% of data blob size is produced, respectively. In Table 3, we calculate the total amount of redundant data generated with each of the configurations using $(qc) + (d/p)$. Referring to the tables, we observe that 51% reliability is achievable with 11.6% redundancy on data for a significant number of bit errors using the proposed model.

TABLE 2
RELIABILITY PREDICTION

| Data Bits ($d$) | Parity Blocks ($p$) | Checksum Blocks ($c$) | Checksum Output Size ($q$) | Bit Errors ($n$) | Reliability |
|---|---|---|---|---|---|
| $8 \times 10^6$ | 2 | 15625 | 16 | 1000 | 0.8811824 |
| $8 \times 10^6$ | 2 | 7813 | 16 | 1000 | 0.8796356 |
| $8 \times 10^6$ | 5 | 15625 | 16 | 1000 | 0.7280075 |
| $8 \times 10^6$ | 5 | 7813 | 16 | 1000 | 0.7232780 |
| $8 \times 10^6$ | 10 | 15625 | 16 | 1000 | 0.5269373 |
| $8 \times 10^6$ | 10 | 7813 | 16 | 1000 | 0.5136633 |
| $8 \times 10^6$ | 20 | 15625 | 16 | 1000 | 0.2647691 |
| $8 \times 10^6$ | 20 | 7813 | 16 | 1000 | 0.2118513 |

TABLE 3
REDUNDANT DATA SIZE

| Data Bits ($d$) | Parity Blocks ($p$) | Checksum Blocks ($c$) | Checksum Output Size ($q$) | Redundant Data Bits |
|---|---|---|---|---|
| $8 \times 10^6$ | 2 | 15625 | 16 | 4250000 (531kB) |
| $8 \times 10^6$ | 2 | 7813 | 16 | 4125008 (516kB) |
| $8 \times 10^6$ | 5 | 15625 | 16 | 1850000 (231kB) |
| $8 \times 10^6$ | 5 | 7813 | 16 | 1725008 (216kB) |
| $8 \times 10^6$ | 10 | 15625 | 16 | 1050000 (131kB) |
| $8 \times 10^6$ | 10 | 7813 | 16 | 925008 (116kB) |
| $8 \times 10^6$ | 20 | 15625 | 16 | 650000 (81kB) |
| $8 \times 10^6$ | 20 | 7813 | 16 | 525008 (66kB) |

### B. PRACTICAL IMPLEMENTATION

In this section, we build the proposed system in real life as a proof of concept. Although simple in design, the software implementation is to serve as a working standard that can be easily re-implemented in any programming language or developed on the initial version.

The presented program "Regen", short for "Redundancy Generator" features its file format with the extension "regen". These files contain the redundant data generated by the program. Table 4 describes the structure of the regen file. The header starts with an ASCII-encoded magic sequence, followed by three 16-bit numbers which are used to infer the structure of its body and the version of the program used to produce the data. Wherever integers are read or written to file, unsigned big-endian encoding should be used.

TABLE 4
REGEN FILE FORMAT

| Section | Length (bytes) | Type | Description |
|---|---|---|---|
| Header | 5 | byte[5] | Magic sequence: "REGEN" |
| Header | 2 | uint16 | Version number |
| Header | 2 | uint16 | Checksum block length in bytes |
| Header | 2 | uint16 | Parity blocks count |
| Body | Variable | byte[] | Checksum data |
| Body | Variable | byte[] | Parity data |

Each of the functionality provided by the program is described in detail below consisting of procedures given in pseudocode.

### 1) GENERATE

The generate procedure has two outputs, a hash file with the SHA256 hash of the archive file and a regen file with the redundant data. The procedure consumes three arguments, archive filename, parity percentage, and checksum block length, in this order.

First, a SHA256 hash is generated from the archive file and saved by the same name as the archive file with a SHA256 extension. The saved hash is used later on to test the integrity of the archive file and to see if a regenerate process is necessary. SHA256 is a highly effective cryptographic hash algorithm that can quickly generate and verify against large archives of data [16].

Next, the redundancy generation process starts by creating the regen file and writing header data with the checksum block length and number of parity blocks. The checksum block count is calculated and we iterate the first set of two loops over each parity block and checksum block. Fletcher-16 checksum is calculated for each of the blocks and written to the regen file sequentially. Once this is done, buffer arrays are initialized to load parity data from the source archive and the output from the XOR parity function. Now we iterate for each chunk of buffer size over every parity block to generate the parity data and proceed to write to the regen file. Once that is completed, we should have a hash file and regen file generated from the provided archive file. The regen file has a fixed-size header that can describe the file's body contents together with the archive file information.

The Fletcher algorithm should use a modulus of 255 and one's complement addition. This ensures optimal performance [17]. In addition, the algorithm has shown a steep degradation in performance when block sizes larger than 256 bytes were used, therefore it is not advised to use larger block sizes [17].

---

**Algorithm 1** Generate on Archive

GENERATE(*afn*, *p*, *cbl*)
1  $af \leftarrow$ OPEN(*afn*)
2  $hash \leftarrow$ SHA256(*af*)
3  $hf \leftarrow$ CREATE(*afn* + ".sha256")
4  $hf$.WRITE(*hash*)
5  $afi \leftarrow$ STAT(*af*)
6  $pb \leftarrow$ ROUND(100 / *p*)
7  $pbl \leftarrow$ FLOOR(*afi*.SIZE() / *pb*)
8  $rf \leftarrow$ CREATE(*afn* + ".regen")
9  $header \leftarrow$ **new** BYTE[11]
10  $header[0:5] \leftarrow$ "REGEN"
11  $header[5:7] \leftarrow 1$
12  $header[7:9] \leftarrow cbl$
13  $header[9:11] \leftarrow pb$
14  $rf$.WRITE(*header*)
15  $cb \leftarrow$ CEIL(*pbl*/*cbl*)
16  $lcbl \leftarrow pbl - ((cb - 1) * cbl)$
17  $cbb \leftarrow$ **new** BYTE[*cbl*]
18  **for** $i = 0 : pb$ **do**
19    **for** $j = 0 : cb$ do
20      $offset \leftarrow pbl + j * cbl$
21      $af$.READ_AT(*cbb*, *offset*)
22      **if** $j == cb - 1$ **then**
23        **if** $lcbl == 0$ **then**
24          **continue**
25        $sum \leftarrow$ FLETCHER16(*cbb*[:*lcbl*])
26      **else**
27        $sum \leftarrow$ FLETCHER16(*cbb*)
28      $rf$.WRITE(*sum*)
29  $pbuff \leftarrow$ **new** BYTE[*pb*]
30  $pobuff \leftarrow$ **new** BYTE[*pbl*]
31  **for** $i = 0 : pb$ **do**
32    $offset \leftarrow i * pbl$
33    $af$.READ_AT(*pbuff*[*i*], *offset*)
34  $pobuff \leftarrow$ XOR_PARITY(*pbuff*)
35  $rf$.WRITE(*pobuff*)

---

### 2) VERIFY

A SHA256 hash is calculated from the archive file and compared with the hex-encoded hash file contents. If the digest does not match, the user is reported about the data corruption so they can use the regenerate feature and restore the integrity.

The process of verifying the data integrity of an archive is quite standard. This allows the user to calculate the hash of the archive file using any cryptographic tool and compare it with the saved hash file.

### 3) REGENERATE

If the archive file is found to have a hash mismatch. The regen file with its redundancy data is used to recover the integrity by finding and fixing any errors in data. The procedure takes only the archive filename as an argument. The regen file is identified by the same name with its extension. Header data is read from the regen file to identify the number of checksum blocks and parity blocks written to the file. Buffer arrays are initialized for checksum data and parity data. Arrays are initialized to store corrected block information and failed block indexes as well.

Now we go through every checksum block in every parity block in sequence. Fletcher-16 is computed for the checksum block, fetching and comparing with the corresponding checksum from the regen file. If a mismatch is detected, we calculate parity for the checksum block contents. Computed parity data is compared with corresponding parity data in the regen file bit by bit and faulty bit indexes are noted. Now we use a function to generate a set of all possible combinations with bad bit indexes, from more values to fewer values. A buffer for correction applied checksum block data is initialized and checksum block data with each combination is tried until the correct checksum is found. If a correct combination is found, we save it in the corrected blocks array. If no correct combination is found, the checksum block index is inserted into the failed blocks array.

At this point, we should have an array of corrected block objects and failed block indexes. We then iterate through the corrected block objects and test if its checksum block index happens to be in the list of failed checksum block indexes. If this is not the case, we write the correction into the archive file in the proper location.

The chances of successful recovery depend on the number of bit errors in the archive file, the number of parity blocks, and the checksum blocks used to generate the redundancy data. When generating bit combinations, a limit of 10 bits and 1023 arrangements could be implemented as a threshold for computation complexity and collision probability.

| Algorithm 2 Regenerate from Redundant Data |
|---|

REGENERATE(*afn*)
1  *af* ← OPEN(*afn*)
2  *afi* ← STAT(af)
3  *rf* ← OPEN(*afn* + ".regen")
4  *header* ← **new** BYTE[11]
5  *rf*.READ(*header*)
6  *cbl* ← UINT16(*header*[7:9])
7  *pb* ← UINT16(*header*[9:11])
8  *pbl* ← FLOOR(*afi*.SIZE() / *pb*)
9  *cb* ← CEIL(*pbl* / *cbl*)
10 *lcbl* ← *pbl* – ((*cb* - 1) * *cbl*)
11 *cbb* ← **new** BYTE[*cbl*]
12 *cbuff* ← **new** BYTE[2]
13 *pbuff* ← **new** BYTE[*pb*][*cbl*]
14 *pobuff* ← **new** BYTE[*cbl*]
15 *rpbuff* ← **new** BYTE[*cbl*]
16 *cbs* ← **new** CBLOCK[]
17 *fbs* ← **new** INT[]
18 **for** *i* = 0 : *pb* **do**
19   **for** *j* = 0 : *cbs* **do**
20     *offset* ← (*i* * *pbl*) + (*j* * *cbl*)
21     *af*.READ_AT(*cbb*, *offset*)
22     *lcb* ← *j* == (*cb* - 1)
23     **if** *lcb* **then**
24       **if** *lcbl* == 0 **then**
25         **continue**
26       *sum* ← FLETCHER16(*cbb*[:*lcbl*])
27     **else**
28       *sum* ← FLETCHER16(*cbb*)
29     *rf*.READ_AT(*cbuff*, 11 + (*i* * (*cb* * 2)) + (*j* * 2))
30     *rsum* ← UINT16(*cbuff*)
31     **if** *sum* != *rsum* **then**
32       *rpoffset* ← 11 + (*cb* * 2 * *pb*)
33       *rf*.READ_AT(*rpbuff*, *rpoffset* + *j* * *cbl*)
34       **for** *k* = 0 : *pb* **do**
35         *af*.READ_AT(*pbuff*[*k*], *k* * *pbl* + *j* * *cbl*)
36       *pobuff* ← XOR_PARITY(*pbuff*)
37       **if** *lcb* **then**
38         *bl* ← *lcbl*
39       **else**
40         *bl* ← *cbl*
41       *bbits* ← **new** INT[]
42       **for** *k* = 0 : *bl* **do**
43         *pbyte* ← *pobuff*[*k*]
44         *rbyte* ← *rpbuff*[*k*]
45         **for** *l* = 0 : 8 **do**
46           *bmask* ← BYTE(1 << *l*)
47           *pbbyte* ← *pbyte* & *bmask*
48           *rbbyte* ← *rbyte* & *bmask*
49           **if** *pbbyte* != *rbbyte* **then**
50             *bbits*.APPEND((*k* * 8) + *l*)
51       *combs* ← GENERATE_COMBS(*bbits*)
52       *cbcbuff* ← **new** BYTE[*cbl*]
53       *found* ← **false**
54       **for** Each *comb* in *combs* **do**
55         *cbcbuff* ← *cbbuff*
56         **for** Each *index* in *comb* **do**
57           *fbyte* ← FLOOR(*index* / 8)
58           *fmask* ← BYTE(1 << (*index* % 8))
59           *cbcbuff*[*fbyte*] ← *cbcbuff*[*fbyte*] ^ *fmask*
60         *sum* ← FLETCHER16(*cbcbuff*)
61         **if** *sum* == UINT16(*cbuff*) **then**
62           *found* ← **true**
63           **break**
64       **if** *found* **then**
65         **if** *lcb* **then**
66           *cblock* ← CBLOCK{*offset*, *j*, *lcbl*, *cbcbuff*}
67         **else**
68           *cblock* ← CBLOCK{*offset*, *j*, *cbl*, *cbcbuff*}
69         *cbs*.APPEND(*cblock*)
70       **else**
71         *fbs*.APPEND(*j*)
72 **for** Each *block* in *cbs* **do**
73   *bfailed* ← **false**
74   **for** Each *i* in *fbs* **do**
75     **if** *i* == *block*.*cb* **then**
76       *bfailed* ← **true**

```
77    if not bfailed then
78        buff ← new BYTE[block.bl]
79        buff ← block.cbcb
80        af.WRITE_AT(buff, block.offset)
```

A Go language implementation of the program is available on GitHub [18]. The program features a command line interface and exported functions that can be integrated with other Go language programs. It is a lightweight program with minimal external dependencies, using the standard library for operations such as generating cryptographic hashes. It can be built on all major desktop operating systems and CPU architectures. It includes sample archive data and unit tests to evaluate the functionality and performance of the program. Refer to the repository for optimized working code.

## V. EXPERIMENTS
### A. BENCHMARKING REGEN

The program is evaluated using computer-generated tests to get an empirical, real-world performance metric in terms of reliability. Benchmarking is a form of controlled experiment that can provide a repeatable method to analyze the performance, scalability, or reliability of a software program [19].

Two types of benchmark tests are carried out. First, we conduct a benchmark to evaluate the recoverability of burst errors, $n$ number of errors distributed randomly across $b$ random locations of sequential bursts. Then, we evaluate the recoverability of a random $n$ number of bit errors distributed spontaneously across the archive file. Several different configurations and archive file sizes are tested with this method.

Benchmarking is implemented as Go language unit tests and is executed using Go Tools. All of the tests presented can be reproduced by cloning the code repository and running Go Tools, specifying the respective benchmark. Refer to the Readme document for instructions on how to run the tests.

Before we compare the results, the nature of predictability for the system should be discussed. The number of bit errors or the configuration used for redundancy alone cannot determine if a malformed archive can be recovered. This is because of the non-deterministic nature of some factors contributing to failure. The alignment and distribution of the error bits greatly influence the results. In the benchmarks, we use computer-generated random bit and burst errors. The benchmark results will not be the same each time it is run due to the randomization function.

The burst error benchmark results given in Table 5 far exceed the bit error benchmarks whereas the bit error benchmark results are seen to align with the theoretical predictions given in Table 2 closely. Reliability for bit error recovery is found to be highly determinable since the prediction model was derived for this scenario. The system is seen to provide high reliability for up to 10 burst errors with a small fraction of redundant data. When testing the limits of burst error recovery, we see that an eighth of the data size could be recovered with a high redundancy configuration. Random bit errors however make it more difficult to recover, this is because it can cause checksum errors in more blocks overall and combinations have to be tried in some of them, leading to higher chances of failure. However, this type of malformation can still be successfully recovered where there are a smaller number of error bits, as seen in Table 6. The amount of parity data used has a huge impact on recoverability, therefore, a parity percentage of 5-50% can be used depending on the required reliability. A checksum block size of 64 bytes is seen to provide optimum performance although it does impact the generate time complexity significantly, therefore, 128-byte checksum blocks are more appropriate for large archive files.

TABLE 5
BURST ERROR BENCHMARKS

| Data Size | Errors ($n$) | Bursts ($b$) | Parity Block Size | Checksum Block Size | Successful Recovery |
|---|---|---|---|---|---|
| 1MB | 1000 | 10 | 10% | 32B | 97/100 (97%) |
| 1MB | 1000 | 10 | 10% | 64B | 98/100 (98%) |
| 1MB | 1000 | 10 | 10% | 128B | 94/100 (94%) |
| 1MB | 1000 | 10 | 5% | 64B | 94/100 (94%) |
| 1MB | 1000 | 20 | 5% | 64B | 75/100 (75%) |
| 1MB | 1000 | 40 | 5% | 64B | 34/100 (34%) |
| 1MB | 10000 | 10 | 5% | 64B | 79/100 (79%) |
| 1MB | $10^5$ | 1 | 10% | 64B | 92/100 (92%) |
| 1MB | $10^5$ | 2 | 10% | 64B | 79/100 (79%) |
| 1MB | $10^6$ | 1 | 50% | 64B | 73/100 (73%) |
| 1MB | $10^6$ | 2 | 50% | 64B | 65/100 (65%) |
| 1GB | $10^6$ | 10 | 10% | 64B | 9/10 (90%) |
| 1GB | $10^6$ | 10 | 5% | 128B | 10/10 (100%) |
| 1GB | $10^6$ | 20 | 5% | 128B | 10/10 (100%) |
| 1GB | $10^7$ | 10 | 10% | 128B | 8/10 (80%) |

TABLE 6
BIT ERROR BENCHMARKS

| Data Size | Errors ($n$) | Parity Block Size | Checksum Block Size | Successful Recovery |
|---|---|---|---|---|
| 1MB | 1000 | 10% | 64B | 49/100 (49%) |
| 1MB | 1000 | 50% | 64B | 91/100 (91%) |
| 1MB | 500 | 10% | 64B | 80/100 (80%) |
| 1MB | 500 | 5% | 64B | 69/100 (69%) |
| 1MB | 250 | 10% | 64B | 93/100 (93%) |
| 1MB | 250 | 5% | 64B | 91/100 (91%) |
| 1GB | $10^6$ | 10% | 64B | 0/10 (0%) |
| 1GB | $10^4$ | 10% | 128B | 10/10 (100%) |
| 1GB | $10^4$ | 5% | 128B | 9/10 (90%) |

### B. SIMULATIONS

Experiments are conducted to demonstrate how our method of data redundancy can be used to provide resilience against different types of integrity loss. The experiments done are by design not too elaborate to allow for replication of the exact procedure. The tools required are, Oracle VirtualBox 7, a compatible host operating system, Ubuntu Desktop 22 LTS, Active@ Disk Editor 23, Free Download Manager 6, and Regen 1.

### 1) DISK SECTOR ERROR

Redundancy systems such as RAID focus on whole disk failures even though it is not the only cause of data corruption in storage disks [20], [21], [22]. A case of latent sector error can make a large archive file unusable. Even a minor error can spoil the entire file where it has a compression or encryption encoding applied. It is referred to as a latent sector error when the correct data in a disk sector is not read and the disk ECC is not able to provide a correction [20]. This can happen in deteriorating hard disks. In this experiment, we use a disk editor to locate the physical sectors of a stored archive file, erase data to simulate sector failures, and test if we can restore the integrity of data using Regen.

To perform the experiment, Ubuntu OS is set up in VirtualBox, a virtualization software. We perform the disk manipulation inside a virtual environment to avoid the possibility of unintended damage to the host filesystem. With the Ubuntu installation, we start by cloning Regen v1.0.4 from GitHub. It should have a sample file "cats.zip" in the root directory which we move to the directory "testdata". Make sure no copies are created, otherwise it would make it more difficult for us to identify the corresponding data for our sample file inside of the virtualized disks. Proceed by generating redundancy on the sample archive with parameters 5% parity and 64-byte checksum block length. We should now have a SHA256 and regen file generated inside the test data directory. Now launch the Active@ Disk Editor program to locate our sample archive data. In Disk Editor, go to find, then insert the first 16 bytes of our sample file in hex, select per block search, enter block size 512 bytes, and click find all. We observed that the ext4 filesystem aligns our data with 512-byte blocks. So, by executing a per-block search, we can significantly reduce the time it takes to identify the location of our data. The data address should be identified and presented in the find results panel. Select the search result and the view will jump to the address of our archive data. In current generation hard disk drives, normally 4 Kibibyte physical sector size is used. Therefore, we will select and erase 4096 bytes of data from the head of the archive. Highlight the data to erase as shown in Fig. 2. Go to edit and select allow edit content. Again, go to edit and select fill block, here we can enter hex value 00 and click ok to convert the selected area into null bytes. Now, click save to write the changes to disk. Once done, the system must be rebooted so that any disk page caching is cleared from primary memory.

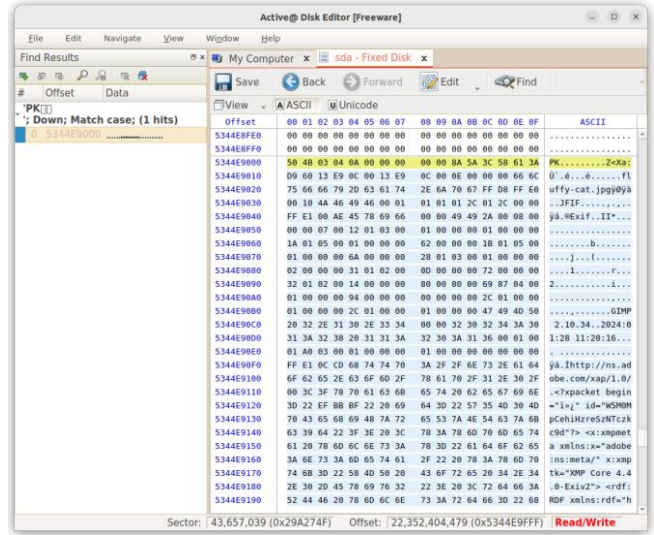

**FIGURE 2.** Data selected in Disk Editor. Physical sectors are highlighted for erasure.

Once the system is rebooted, we use Regen to verify the data integrity by comparing it against the saved SHA256 file. It should now report that the archive file is corrupt. We then run regenerate on the archive file. Regen will use the saved redundancy data to restore the archive file to its original form. The result can be verified by running the verify command again.

We then proceed to test sector errors on a larger, 1GB file. Use wget to download ubuntu-20.10-live-server-amd64.iso into the test data directory from Ubuntu archives. Generate redundancy with parameters 5% parity and 128-byte checksum block length. Now go to Disk Editor and search for the archive data, similarly. Once the data sector is located, select the first 40960 bytes and erase. Reboot the system, use regenerate, and validate the result by running verify.

In this experiment, we erased physical disk sectors from the ext4 filesystem to simulate latent sector errors and found out that the manipulation remains dormant and uncorrected by the operating system at the time when it is accessed. Generated partial redundancy on the archives was used to successfully recover the sample files in both of the test cases. This confirms the effectiveness of the proposed system for the recovery of minor errors to archive data. Particularly large archive files which will be lost entirely from the smallest error.

### 2) TRANSMISSION ERROR

Data transfers are mostly done over protocols utilizing TCP in the transport layer. This means error detection and retransmission capability are provided by the networking layer itself [23]. Nevertheless, issues such as system failures or user errors can still lead to faulty data being delivered. Moreover, alternative UDP-based data transfer protocols exist such as Trivial File Transfer Protocol, UDP-based Data Transfer Protocol, and Micro Transport Protocol [24], [25], [26]. This can potentially reduce the reliability of transmitted data. Therefore, additional error correction mechanisms could be utilized.

In this experiment, we consider the possibility of using partial redundancy data to reconstruct a large file download that fails integrity verification due to a minor error. In this scenario, the download provider will make available both the SHA256 hash and regen file.

To simulate transmission failure, we will interrupt a download at 99% progress using Free Download Manager. Find the link for ubuntu-20.10-live-server-amd64.iso and add it to the download manager. Watch the progress and select pause when the download is at 99% as shown in Fig. 3. Open the progress view and it will show missing data blocks and their locations. The empty blocks are scattered across the file because the download manager uses multiple simultaneous connections to stream data in parallel. The incomplete file should be in the Downloads directory with a "fdmdownload" extension. Copy this into the test data directory with the extension removed. We should have the required SHA256 and regen file in the test data directory from the previous experiment. Start by running the verify command with Regen. It will show inconsistent hashes. Now we execute the regenerate command. This will reconstruct our failed download file using the generated redundant data with configurations of 5% parity and 128-byte checksum blocks. When done, use verify again and it will now show consistent hashes.

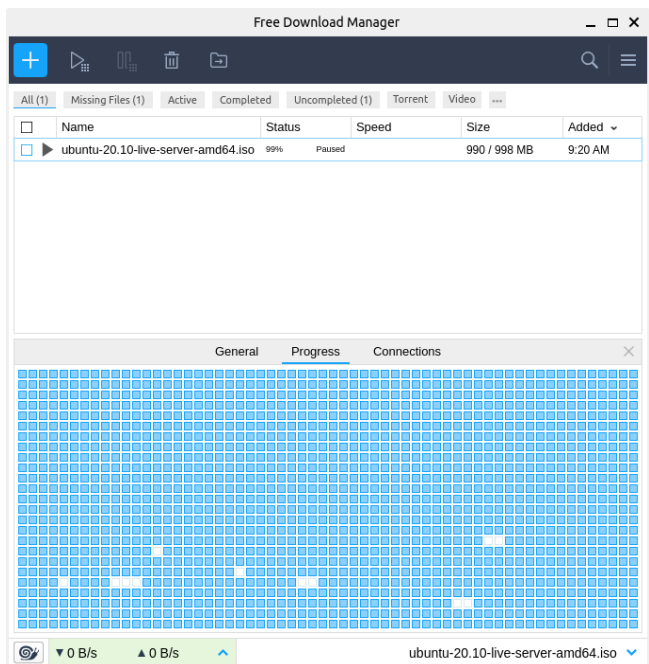

**FIGURE 3.** Download interrupted at 99%. Empty areas in the block grid indicate missing data.

## VI. CONCLUSION

In this paper, we have presented a novel design for a data redundancy system featuring concepts such as multiple layers of abstraction for parity blocks and checksum blocks, the use of fast arithmetic checksums in combination with parity data and a brute force combination search method, discussing an appropriate checksum configuration to balance computing power requirement and collision probability. In the experiments, we demonstrated that the produced software tool can effectively increase the reliability of large archive files for long-term preservation in hard disk drives by providing resistance to faults such as latent sector errors. The system performed robustly in more experiments as well, indicating that it may have a broad scope of application.